\documentclass[notitlepage,showkeys,twocolumn,11pt,times,superscriptaddress, a4paper]{revtex4-2}
\usepackage{stix}
\usepackage[T1]{fontenc}
\usepackage[english]{babel}
\usepackage{graphicx}
\usepackage{bm} 
\usepackage[colorlinks=true,
            linkcolor=blue,
            urlcolor=blue,
            citecolor=blue]{hyperref}
\usepackage[mathlines]{lineno} 
\usepackage{amsmath,mathtools}
\usepackage{physics}
\usepackage{lipsum}
\usepackage{booktabs}
\usepackage{geometry}
\DeclareMathOperator{\sgn}{sgn}

\begin{document}
    \title{Adiabatic confinement and escape in an open billiard}

    \author{Matheus Rolim Sales}
    \email{rolim.sales.m@gmail.com}
    \affiliation{São Paulo State University (UNESP), Institute of Geosciences and Exact Sciences, 13506-900, Rio Claro, SP, Brazil}

    \author{Leonardo Costa de Souza}
    \affiliation{Institute of Physics, University of São Paulo, 05315-970, São Paulo, SP, Brazil}
    
    \author{Ricardo Luiz Viana}
    \affiliation{Federal University of Paraná, Department of Physics, Interdisciplinary Center for Science, Technology and Innovation, Center for Modeling and Scientific Computing, 81531-980, Curitiba, PR, Brazil}

    \author{Edson Denis Leonel}
    \affiliation{São Paulo State University (UNESP), Institute of Geosciences and Exact Sciences, 13506-900, Rio Claro, SP, Brazil}

    \author{Iberê Luiz Caldas}
    \affiliation{Institute of Physics, University of São Paulo, 05315-970, São Paulo, SP, Brazil}

    \date{\today}

    \begin{abstract}
        We introduce a stationary, field-free billiard that reproduces the adiabatic confinement mechanism of a magnetic mirror through boundary geometry alone. Rapid transverse reflections generate an approximately conserved action that plays the role of the magnetic moment, producing an effective longitudinal barrier and a nominal loss cone for escape. For the half-width profile considered here, confinement is controlled by a competition between mirror strength and adiabaticity: shortening the profile increases the mirror ratio but also reduces the accuracy of the adiabatic approximation. As a result, the escaping fraction is nonmonotonic and confinement is strongest at an intermediate profile length. We characterize the associated phase-space structure, escape-time statistics, and escape basins, revealing long-lived transient motion and a strong sensitivity of the final escape direction to the initial condition that persists as the uncertainty in the initial condition is reduced.
    \end{abstract}

    \maketitle

    \section{Introduction}

    Many dynamical systems exhibit motions occurring on widely separated time scales. When a fast degree of freedom evolves while the parameters governing its motion vary sufficiently slowly, an associated action can remain approximately conserved, giving rise to an adiabatic invariant~\cite{Gardner1959,Wells2007,Brown1987}. Such invariants provide an effective reduction of the dynamics and can strongly constrain transport even when no exact conservation law is present. A particularly important example occurs in the motion of charged particles in a slowly varying magnetic field, for which the magnetic moment associated with the fast gyromotion is approximately conserved~\cite{Northrop1963}. As a particle moves into a region of increasing magnetic field strength, conservation of this quantity redistributes kinetic energy from the motion along the field to the perpendicular motion and can eventually reverse the parallel velocity. This mechanism forms the basis of magnetic-mirror confinement~\cite{Northrop1960,Schmidt1979,Chen2015}.

    The same mechanism of confinement through an adiabatic invariant can arise in systems where the separation of time scales is generated geometrically rather than by a slowly varying field and billiards provide a particularly simple class of Hamiltonian systems in which the dynamics is directly connected to geometry~\cite{Tabachnikov2005,Chernov2006}. In their two-dimensional formulation, a point-like particle moves freely within a planar domain and undergoes specular reflections at its boundary. Despite this simple construction, billiards can exhibit regular, chaotic, or mixed dynamics depending on the geometry of the boundary~\cite{Sinai1970,Bunimovich1979,Chernov2006,Bunimovich2018}. Dispersing boundaries generate the hyperbolicity of the Sinai billiard~\cite{Sinai1970}, while Bunimovich showed that chaotic dynamics can also arise in billiards containing focusing components through the defocusing mechanism~\cite{Bunimovich1979,Bunimovich2018}. More generally, changes in boundary geometry can produce qualitatively different phase-space structures, ranging from predominantly regular motion to strongly chaotic dynamics~\cite{Robnik1983, Bunimovich2001, Altmann2006}. Billiards therefore provide a useful framework for isolating the consequences of geometry for transport of chaotic orbits~\cite{Artuso1996, Altmann2006, Lozej2018, Diaz2019, Araujolima2024, Carmo2024, Carmo2026} and for escape in open systems~\cite{Dettmann2009, Custodio2011, Hansen2018, Sales2024, Borin2026, Haerter2026}. We also refer the reader to the review by Altmann et al.~\cite{Altmann2013} and references therein, where the authors discuss transient chaos in open systems in general.

    This motivates considering a billiard in the form of a slowly varying channel, in which rapid transverse reflections occur while the longitudinal motion evolves on a comparatively slow scale. As the channel narrows, approximate conservation of the transverse action increases the transverse component of the particle velocity at the expense of its longitudinal component, eventually producing a reversal of the longitudinal motion. A suitably shaped billiard can therefore reproduce the essential adiabatic mechanism of a magnetic mirror without the presence of a magnetic field.

    Adiabatic invariance has previously been explored in several billiard-related settings. Samec et al.~introduced a geometric mirror in which charged particles are confined within a slowly narrowing cylindrical region by an adiabatic invariant associated with the transverse motion~\cite{Samec1975}. In that system, however, reflection at the boundary is produced by a localized magnetic field. Adiabatic invariants have also been studied in billiards whose boundaries vary slowly in time, including both chaotic billiards and nonautonomous billiards exhibiting Fermi acceleration~\cite{Brown1987,Gelfreich2012}. A different class is provided by magnetic billiards~\cite{Robnik1985, Berglund1996, Altmann2006}, where charged particles move under a magnetic field while also undergoing specular reflections at a boundary. In this setting, adiabatic invariants have been derived for the resulting slow-fast dynamics, while their destruction has been shown to generate chaotic motion~\cite{Neishtadt2012,Artemyev2015}. These studies establish several connections between adiabatic invariance and billiard dynamics, but differ from a stationary, field-free billiard in which both reflection and confinement arise solely from the geometry of the boundary.

    In this work, we introduce a stationary, field-free billiard designed to reproduce magnetic-mirror confinement and investigate the validity and breakdown of its adiabatic description. We first establish the correspondence between the magnetic moment of the magnetic mirror and the transverse action of the billiard, and derive the associated turning-point and loss-cone conditions. We then examine the validity of this description as the geometry of the billiard is varied. Beyond the adiabatic approximation, we investigate how the phase-space structure affects confinement and escape, with particular attention to the fraction of escaping trajectories, the number of collisions they undergo before escaping, and the dependence of the final escape direction on the initial condition. Finally, we characterize the sensitivity of the escape basins through the uncertainty exponent.

    This paper is organized as follows. In Sec.~\ref{sec:model}, we introduce the magnetic mirror and the corresponding billiard model, derive the adiabatic description, and discuss its range of validity. In Sec.~\ref{sec:phase}, we examine the phase-space structure and confinement properties of the billiard. In Sec.~\ref{sec:escape}, we investigate the escaping fraction and escape-time statistics, while in Sec.~\ref{sec:basins}, we analyze the escape basins and their sensitivity to initial conditions. Finally, in Sec.~\ref{sec:conclusions}, we summarize our main conclusions.

    \section{The billiard}
    \label{sec:model}

    In this section, we construct a billiard that confines particles through the same dynamical mechanism responsible for magnetic-mirror confinement. The essential ingredient is an approximate adiabatic invariant associated with a fast periodic motion, whose slow variation through the system produces an effective barrier for the remaining degree of freedom. We first introduce the magnetic mirror and its loss cone, then define the billiard and establish the correspondence between the two. A more detailed derivation of the magnetic-mirror dynamics is given in the Supplemental Material.

    \subsection{The magnetic mirror}
    \label{sec:mirror}

\begin{figure}[t!]
        \centering
        \includegraphics[width=\linewidth]{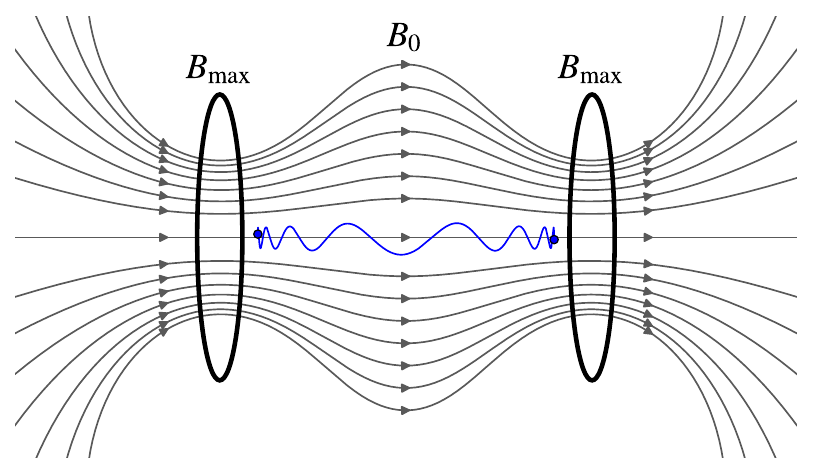}
        \caption{Magnetic mirror formed by two coaxial coils, seen in a plane through the common axis. The coils sit at $z = \pm 1.4a$, where $a$ is their radius, and the rings drawn at those positions represent their cross sections. The grey curves are magnetic field lines, obtained as contours of the magnetic flux of the pair, and the arrows indicate the direction of $\vb{B}$. The field is weakest on the midplane, where it is $B_0$, and increases towards the two throats. The blue curve is a charged-particle trajectory launched from the midplane. A single pass between its two turning points is shown, with the turning points marked by blue dots.}
        \label{fig:magneticmirror}
    \end{figure}

    A magnetic mirror is a configuration in which the magnetic field strength has a minimum at the center of a region and increases towards its two ends. In the laboratory, such a configuration can be produced by two coaxial coils separated along their common axis~\cite{Roth1965,Jiang2017}, as shown in Fig.~\ref{fig:magneticmirror}. Additional coils can be placed between the mirror coils to produce a more uniform magnetic field in the central region~\cite{Laurence1966}. The field is weakest on the midplane and strongest near the coils, which form the two throats of the mirror. Charged particles can therefore be confined between the two regions of strong field without material walls. The same magnetic-mirror mechanism occurs naturally in the Earth's magnetosphere, where charged particles can become trapped in the geomagnetic field and undergo repeated bounce motion between conjugate mirror points in the Northern and Southern Hemispheres, contributing to the confinement of particles in the Van Allen radiation belts~\cite{Northrop1960,UkhorskiySitnov2013,Ripoll2020}.

    The confinement relies on the approximate conservation of the first adiabatic invariant. When the magnetic field varies slowly over a Larmor radius, the particle motion separates into a fast gyromotion about a field line and slow motion of the guiding center along it~\cite{Northrop1963,Schmidt1979,Chen2015}. The action associated with the gyromotion is the magnetic moment,
    \begin{equation*}
        \mu = \frac{\tfrac{1}{2}m v_\perp^2}{B},
    \end{equation*}
    which remains approximately constant under the adiabatic approximation~\cite{Schmidt1979,Chen2015}. As the particle moves into a region of stronger field, conservation of $\mu$ transfers kinetic energy from the motion parallel to the field into the gyromotion. The parallel motion can therefore be reflected before reaching the region of strongest field, providing confinement between the two throats. The full derivation of this effective mirror force and the conditions under which the adiabatic approximation applies is given in the Supplemental Material.

    The resulting loss-cone criterion can be obtained by considering a particle launched from the midplane, where the field strength is $B_0$. Let $\theta$ be the pitch angle between the velocity and the local magnetic field. At a turning point, the parallel velocity vanishes, and conservation of $\mu$ gives
    \begin{equation*}
        B_{\mathrm{turn}} = \frac{B_0}{\sin^2\theta}.
    \end{equation*}
    If $B_{\max}$ denotes the largest field strength accessible to the particle, reflection occurs only when $B_{\mathrm{turn}} \leq B_{\max}$. Defining the magnetic mirror ratio as
    \begin{equation*}
        R_{\rm mag} = \frac{B_{\max}}{B_0},
    \end{equation*}
    the particles that are not reflected satisfy
    \begin{equation}
        \label{eq:coneB}
        \sin^2\theta < \frac{1}{R_{\rm mag}}.
    \end{equation}
    These directions form the loss cone in velocity space~\cite{Schmidt1979,Chen2015}. Its boundary depends on the mirror ratio but not on the particle speed. This is the central feature that motivates the billiard construction: a spatially varying system can produce a velocity-dependent confinement criterion through the approximate conservation of an action associated with a fast periodic motion.

    Figure~\ref{fig:magneticmirror} illustrates the accuracy of this adiabatic description for the magnetic configuration considered here. The trajectory is obtained by directly integrating $m\,\dv*{\vb{v}}{t} = q\,\vb{v}\cross\vb{B}$ in the field of the two coils, without imposing conservation of $\mu$. For the configuration shown, $R_{\rm mag} = 2.64$, giving $\theta_c = 0.66\,\mathrm{rad}$, while the particle is launched at $\theta = 0.70\,\mathrm{rad}$, just outside the loss cone. The particle turns where the field reaches $2.392\,B_0$, compared with the adiabatic prediction $B_0/\sin^2\theta = 2.410\,B_0$, a difference of less than one percent for a gyroradius of $0.12a$. The close agreement confirms that the adiabatic description is accurate for this configuration.


    \subsection{Billiard geometry and collision map}
    \label{sec:billiard}

    We now construct the billiard counterpart of the magnetic mirror. We replace the spatial variation of the magnetic field by the geometry of a channel whose width varies along the longitudinal direction. A particle bouncing between two walls has a fast transverse motion and a slower longitudinal motion, while the wall separation varies along the channel. Narrowing the channel increases the frequency of the transverse motion, playing the role of an increasing magnetic field strength in the mirror. The correspondence between these two systems and the resulting confinement criterion are developed in Sec.~\ref{sec:adiabatic}. We first define the billiard and its collision map.

    A billiard is the free motion of a point particle of mass $m$ inside a region $\mathcal{D}$ of the plane, interrupted by elastic specular reflections at the boundary $\partial\mathcal{D}$~\cite{Tabachnikov2005,Chernov2006}. The boundary is static, so the speed $v = \abs*{\vb{v}}$ is conserved. It therefore only fixes the time scale, and we set $v = 1$ and $m = 1$. The region considered here is a channel that narrows towards both ends,
    \begin{equation*}
        \mathcal{D} = \left\{ (x,y) : \abs*{y} \leq h(x), \quad \abs*{x} \leq x_m \right\},
    \end{equation*}
    where the half-width $h(x) > 0$ is a smooth, even function with a single maximum at $x = 0$. The particle moves freely in the interior, $\abs*{y} < h(x)$, and undergoes specular reflections at the walls $y = \pm h(x)$. The two end segments $x = \pm x_m$, with $\abs*{y} \leq h(x_m)$, are apertures rather than reflecting boundaries. A trajectory reaching either aperture therefore escapes, making the billiard an open system. The set of initial conditions whose trajectories never reach an aperture constitutes the nonescaping set.

    Between successive collisions, the trajectory is a straight segment. The dynamics can therefore be represented by a collision map that takes the state immediately after one reflection to the state immediately after the next. Let $(x,y,v_x,v_y)$ denote the state immediately after a reflection. Until the next collision, the particle's trajectory is given by
    \begin{equation*}
        x(\tau) = x + v_x\tau, \qquad y(\tau) = y + v_y\tau.
    \end{equation*}
    The next collision with either wall is determined by the first positive root of
    \begin{equation}
        \label{eq:wallfun}
        g(\tau) = \abs*{y(\tau)} - h\!\left(x(\tau)\right) = 0.
    \end{equation}
    The search for the root is restricted to times before the trajectory reaches an aperture. For $v_x \neq 0$, the corresponding exit time is
    \begin{equation*}
        \tau_{\mathrm{exit}} = \frac{\sgn(v_x)x_m - x}{v_x}.
    \end{equation*}
    If the aperture is reached before $g(\tau)$ vanishes, the trajectory escapes and the map terminates.

    At a collision, the velocity is reflected specularly about the local tangent to the wall. For the upper wall, the boundary is defined by $\Phi(x,y) = y - h(x) = 0$, whose gradient
    \begin{equation*}
        \label{eq:gradphi}
        \grad\Phi = \left(-h'(x),\,1\right)
    \end{equation*}
    points outward from the billiard. For the lower wall, the corresponding outward normal is $\left(-h'(x),-1\right)$. Both cases can therefore be written as
    \begin{equation*}
        \label{eq:normal}
        \vb{n} = \left(-h'(x),\,\sgn(y)\right), \qquad \abs*{\vb{n}}^2 = 1 + h'(x)^2,
    \end{equation*}
    and the post-collision velocity is
    \begin{equation*}
        \vb{v}\,' = \vb{v} - 2\frac{\vb{v}\cdot\vb{n}}{\abs*{\vb{n}}^2}\vb{n}.
    \end{equation*}

    Introducing a collision index $n$, let $\tau_{n+1}$ be the first positive root of Eq.~\eqref{eq:wallfun} obtained by implicitly solving
    \begin{equation*}
        \abs*{y_n+v_{y,n}\tau_{n+1}}=h\!\left(x_n+v_{x,n}\tau_{n+1}\right).
    \end{equation*}
    Provided that this collision occurs before an aperture is reached, the collision map is
    \begin{equation}
        \label{eq:collisionmap}
        \begin{aligned}
            x_{n+1} &= x_n + v_{x,n}\tau_{n+1}, \\
            y_{n+1} &= y_n + v_{y,n}\tau_{n+1}, \\
            \begin{pmatrix} v_{x,n+1} \\ v_{y,n+1} \end{pmatrix}
            &= \frac{1}{1+q_{n+1}^2}
            \begin{pmatrix}
                1-q_{n+1}^2 & 2s_{n+1}q_{n+1} \\
                2s_{n+1}q_{n+1} & q_{n+1}^2-1
            \end{pmatrix}
            \begin{pmatrix} v_{x,n} \\ v_{y,n} \end{pmatrix},
        \end{aligned}
    \end{equation}
    where $q_{n+1} = h'(x_{n+1})$ and $s_{n+1} = \sgn(y_{n+1})$. At each collision, the position is constrained to the boundary, while the fixed speed removes one further degree of freedom from $(x,y,v_x,v_y)$, so the collision map is two dimensional. The continuous trajectory is recovered by concatenating the free-flight segments generated by this map.

    For the half-width we adopt the Lorentzian profile
    \begin{equation}
        \label{eq:profile}
        h(x) = \frac{h_0}{1 + u^2}, \qquad u \equiv \frac{x}{L},
    \end{equation}
    where $h_0 = h(0)$ is the half-width at the center and $L$ sets the length scale over which the channel narrows, with $h(\pm L) = h_0/2$. We use $h_0$ as the unit of length and take $h_0 = 1$ and $x_m = 10$ in all simulations, leaving $L$ as the only parameter of the profile. The profile decreases monotonically with $\abs*{x}$ and its curvature changes sign at
    \begin{equation*}
        x = \pm\frac{L}{\sqrt{3}},
    \end{equation*}
    where $h''(x) = 0$. These points also correspond to the extrema of the wall slope, giving
    \begin{equation}
        \label{eq:maxslope}
        \max_x\abs*{h'(x)} = \frac{3\sqrt{3}}{8}\frac{h_0}{L}.
    \end{equation}
    The walls are concave towards the interior in the central region $\abs*{x} < L/\sqrt{3}$ and convex in the outer region $\abs*{x} > L/\sqrt{3}$. The corresponding focusing and dispersing segments provide the geometric mechanism for the coexistence of regular and chaotic motion and hence for a mixed phase space~\cite{Sinai1970, Bunimovich1979, Bunimovich2018}.

    \begin{figure*}[t!]
        \centering
        \includegraphics[width=\linewidth]{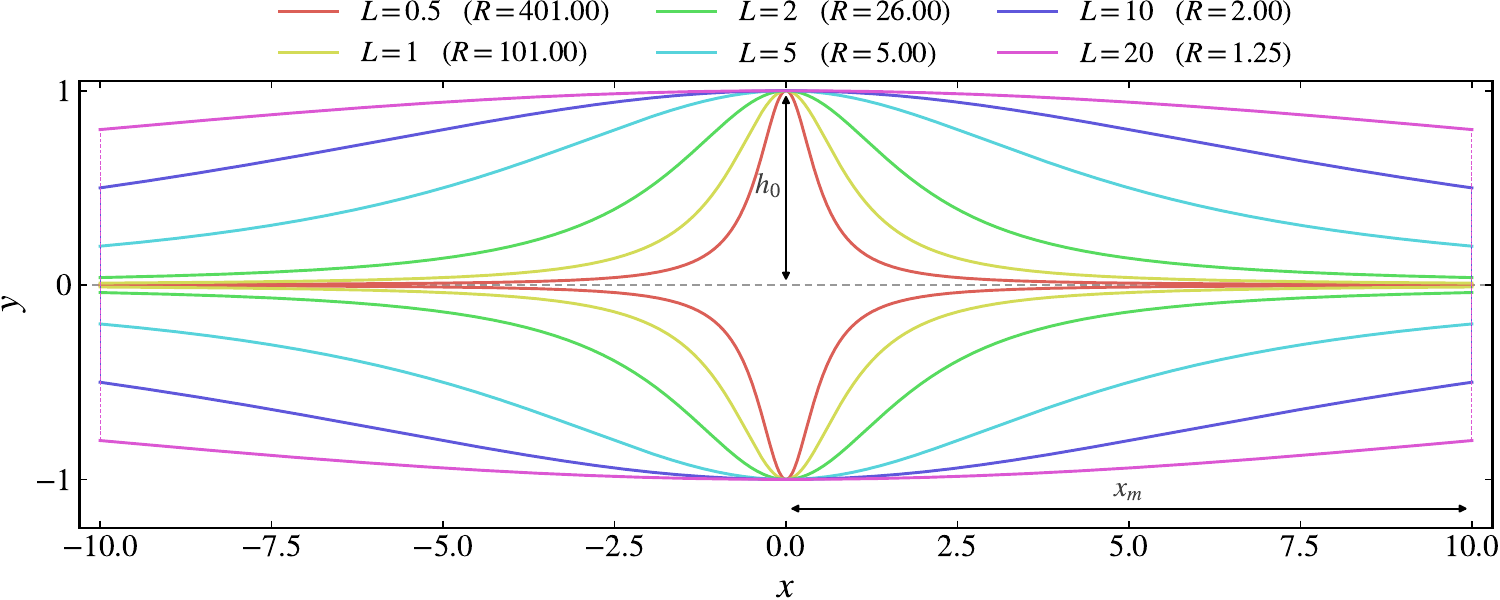}
        \caption{Boundary of the billiard, Eq.~\eqref{eq:profile}, for $h_0 = 1$, $x_m = 10$, and several values of the profile length $L$. All profiles pass through $(0,\pm h_0)$, so $L$ controls the narrowing of the flanks and hence the half-width at the apertures, $h(x_m) = h_0/R$. The dashed vertical lines mark the apertures at $x = \pm x_m$, where the channel is open across $\abs*{y} < h(x_m)$ and particles escape.}
        \label{fig:boundary}
    \end{figure*}

    The quantity that plays the role of the magnetic mirror ratio is the ratio between the maximum and minimum half-widths accessible to the particle,
    \begin{equation}
        \label{eq:R}
        R \equiv \frac{h(0)}{h(x_m)} = 1 + \left(\frac{x_m}{L}\right)^2.
    \end{equation}
    At fixed $h_0$ and $x_m$, this ratio is controlled solely by $L$ and decreases monotonically from $R \to \infty$ as $L \to 0$ to $R \to 1$ as $L \to \infty$. Thus, small $L$ produces a sharply narrowing channel with a large geometric mirror ratio, whereas large $L$ produces a nearly uniform channel with $R$ close to unity. Figure~\ref{fig:boundary} shows the boundary of the billiard for $h_0 = 1$, $x_m = 10$, and different values of $L$, for which $R$ ranges from $R = 401$ at $L = 0.5$ to $R = 1.25$ at $L = 20$.

    \subsection{Adiabatic confinement and loss cone}
    \label{sec:adiabatic}

    We now turn to the adiabatic description of the billiard and derive the corresponding confinement criterion. Suppose that, during one transverse oscillation, the particle moves only a small distance along $x$. The coordinate $x$ can then be regarded as approximately fixed over that oscillation, so that the particle moves between two locally parallel walls at $y = \pm h(x)$. For fixed $x$, the transverse motion is described by $y$ and its conjugate momentum $p_y$, while the half-width $h(x)$ varies slowly as the particle moves along the channel. The action associated with this transverse motion is
    \begin{equation*}
        J_\perp = \frac{1}{2\pi}\oint p_y\,\dd y.
    \end{equation*}
    Let $v_\perp = \abs*{v_y}$ denote the magnitude of the transverse velocity. During the motion from the lower wall to the upper wall, $p_y = m v_\perp$, whereas on the return trajectory $p_y = -m v_\perp$. The action over one complete transverse oscillation is therefore
    \begin{equation*}
        \begin{aligned}
            J_\perp &= \frac{1}{2\pi}\left[\int_{-h}^{h}m v_\perp\,\dd y + \int_{h}^{-h}(-m v_\perp)\,\dd y\right] \\
            &= \frac{2}{\pi}m v_\perp h(x).
        \end{aligned}
    \end{equation*}
    If $h(x)$ changes little during each oscillation, the standard adiabatic approximation implies that $J_\perp$ remains approximately constant~\cite{Gardner1959,Wells2007,Neishtadt2019}. In the exact billiard, successive collisions occur with walls whose orientation changes along the channel and produce small changes in $v_\perp$, so this conservation is only approximate. The role of $J_\perp$ is analogous to that of the magnetic moment $\mu$ in a slowly varying magnetic field~\cite{Northrop1963,Schmidt1979,Chen2015}.

    Taking $x_0$ as a reference position along the channel, approximate conservation of $J_\perp$ gives
    \begin{equation}
        \label{eq:vperpx}
        v_\perp(x) = v_\perp(x_0)\frac{h(x_0)}{h(x)}.
    \end{equation}
    The corresponding transverse kinetic energy is therefore
    \begin{equation}
        \label{eq:kperp}
        K_\perp(x) = \frac{1}{2}m v_\perp^2(x_0)\left[\frac{h(x_0)}{h(x)}\right]^2.
    \end{equation}
    Since the total kinetic energy is conserved, the increase of $K_\perp$ as the channel narrows occurs at the expense of the longitudinal kinetic energy. The transverse motion therefore generates an effective potential for the longitudinal dynamics proportional to $h(x)^{-2}$. A particle moving towards either aperture is progressively decelerated and, if the channel becomes sufficiently narrow, can reach $v_\parallel = 0$ and reverse its direction. This reflection is not produced by a collision with a wall at the turning point, since the walls remain a finite distance apart there. Instead, it results from the cumulative effect of many collisions with walls inclined relative to the longitudinal direction. Each collision transfers a small amount of longitudinal momentum into transverse motion, and the adiabatic invariant describes the resulting cumulative transfer through the approximate conservation of $v_\perp h$.

    The magnetic mirror has the same dynamical structure. Conservation of the magnetic moment gives a perpendicular energy $\mu B(x)$, which acts as an effective potential for the motion along the magnetic field. The quantitative correspondence between the two effective potentials is therefore
    \begin{equation*}
        \frac{B(x)}{B_0} \longleftrightarrow \left[\frac{h_0}{h(x)}\right]^2.
    \end{equation*}
    Thus, increasing the magnetic-field strength in the magnetic mirror corresponds to narrowing the channel in the billiard. The two systems share the same adiabatic confinement mechanism, although the underlying fast motions are different: gyromotion in the magnetic mirror and transverse bouncing in the billiard.

    In the billiard, the pitch angle $\theta$ is the angle between the velocity and the channel axis, so that $v_\parallel = v\cos\theta$ and $v_\perp = v\abs*{\sin\theta}$. For a particle launched from the midplane $x = 0$, where $h(0) = h_0$, with pitch angle $\theta$, Eq.~\eqref{eq:vperpx} gives
    \begin{equation*}
        v_\perp(x) = v\sin\theta\,\frac{h_0}{h(x)}.
    \end{equation*}
    Its nominal turning point $x_t$ is defined by $v_\parallel = 0$, or equivalently $v_\perp = v$, which gives
    \begin{equation}
        \label{eq:turning}
        h(x_t) = h_0\sin\theta.
    \end{equation}
    For the profile in Eq.~\eqref{eq:profile}, this condition gives
    \begin{equation}
        \label{eq:xt}
        x_t = \pm L\sqrt{\frac{1}{\sin\theta} - 1}.
    \end{equation}
    Since $h$ decreases monotonically with $\abs*{x}$, the nominal turning point lies inside the billiard when $h_0\sin\theta > h(x_m) = h_0/R$. Thus, particles satisfying
    \begin{equation}
        \label{eq:cone}
        \sin\theta < \sin\theta_c = \frac{1}{R}
    \end{equation}
    have no turning point before reaching an aperture and escape from the billiard. Equation~\eqref{eq:cone} therefore defines the loss cone, with $\theta_c$ as its critical pitch angle. As in the magnetic mirror, this criterion depends on the direction of the velocity and not on its magnitude. Both $J_\perp$ and $v$ are proportional to the speed, which cancels from the turning condition $v_\perp = v$. For a particle launched away from the midplane, at $x = x_0$, the same argument gives
    \begin{equation*}
        \sin\theta < \frac{h(x_m)}{h(x_0)}.
    \end{equation*}
    The corresponding effective mirror ratio is $h(x_0)/h(x_m)$, which is smaller than $R$ for $x_0 \neq 0$. A particle launched away from the center therefore experiences a weaker mirror and a wider loss cone.

    \begin{figure}[t!]
        \centering
        \includegraphics[width=\linewidth]{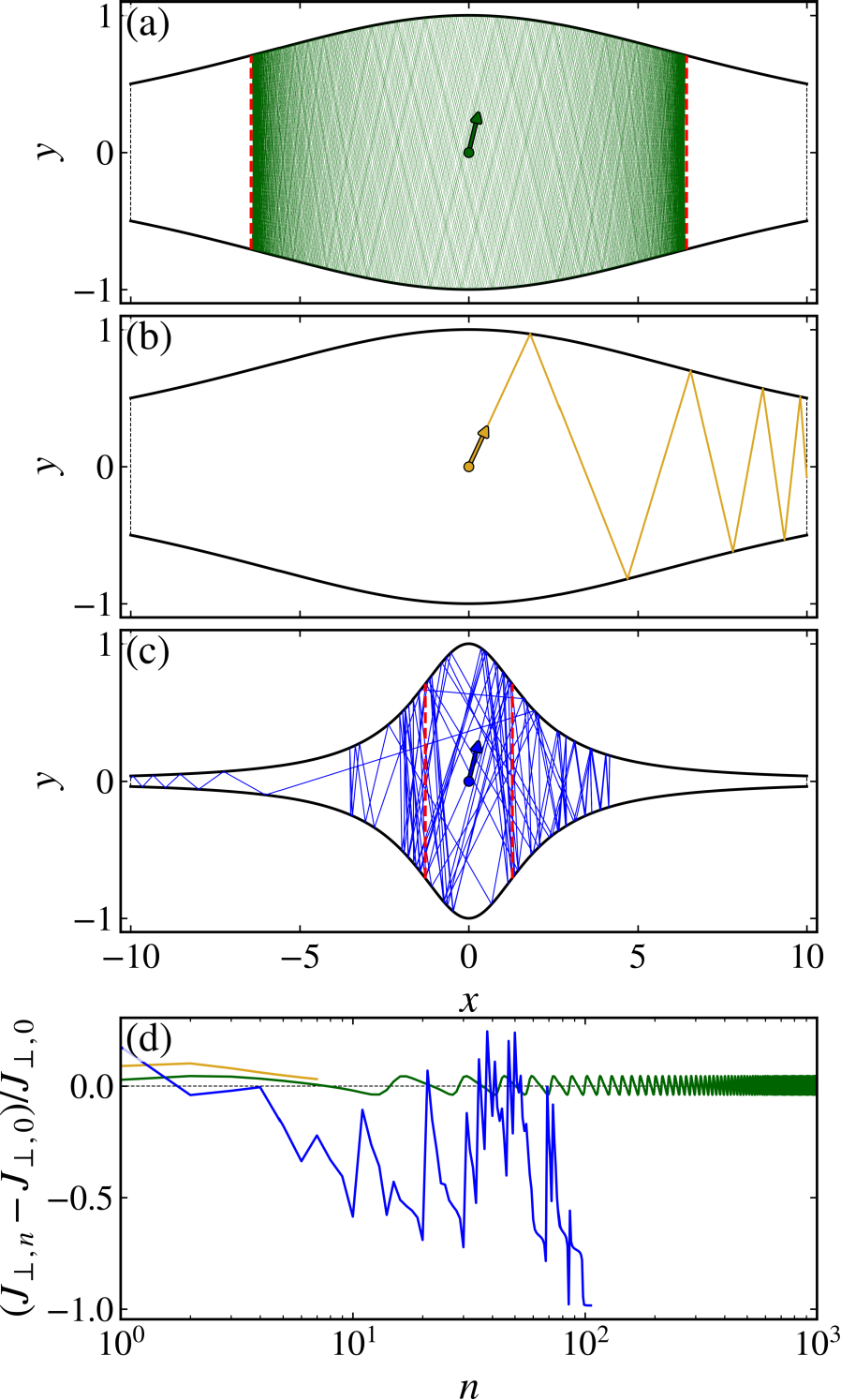}
        \caption{Trajectories of the billiard for $h_0 = 1$ and $x_m = 10$, launched from the origin with unit speed. The dot marks the initial condition and the arrow the initial direction. (a) $L = 10$, $R = 2$ and $\theta = \pi/4\,\mathrm{rad}$. (b) $L = 10$ and $\theta = 0.489\,\mathrm{rad}$, inside a loss cone of critical angle $\theta_c = \pi/6\,\mathrm{rad}$. (c) $L = 2$, $R = 26$ and $\theta = \pi/4\,\mathrm{rad}$, outside a loss cone of critical angle $\theta_c = 0.0385\,\mathrm{rad}$. All angles are in radians. The solid curves are the walls $y = \pm h(x)$ and the dashed segments are the apertures at $x = \pm x_m$. The vertical red dashed lines indicate the nominal turning points predicted by Eq.~\eqref{eq:xt}, at $x_t = \pm 6.44$ in (a) and $x_t = \pm 1.29$ in (c). In (b), Eq.~\eqref{eq:xt} gives $x_t = 10.63$, beyond the aperture, so no turning-point lines are shown. (d) Relative variation $(J_{\perp,n} - J_{\perp,0})/J_{\perp,0}$ of the transverse action as a function of the collision number $n$ for the trajectories in (a)--(c), with the colors corresponding to those of the respective trajectories. The vertical scale in (a)--(c) is stretched with respect to the horizontal one.}
        \label{fig:collisions}
    \end{figure}

    Figure~\ref{fig:collisions}(a) shows a trajectory for $L = 10$ and $\theta = \pi/4\,\mathrm{rad}$, above the critical angle $\theta_c = \pi/6\,\mathrm{rad}$. The nominal turning points from Eq.~\eqref{eq:xt} are $x_t = \pm 6.44$, indicated by the vertical red dashed lines. As the particle approaches either turning point, the longitudinal distance between successive collisions decreases and the motion becomes increasingly transverse. The longitudinal velocity eventually reverses near the predicted location, and the particle remains confined between the two nominal turning points. Figure~\ref{fig:collisions}(b) keeps the same profile but lowers the launch angle to $\theta = 0.489\,\mathrm{rad}$, inside the loss cone. Equation~\eqref{eq:xt} then gives $x_t = 10.63$, slightly beyond the aperture at $x_m = 10$. The particle consequently reaches the right aperture before a turning point can occur and escapes. The two trajectories differ only in their initial direction, illustrating the loss-cone criterion of Eq.~\eqref{eq:cone}. At the critical angle, the nominal turning point coincides with the aperture. Figure~\ref{fig:collisions}(c) illustrates the breakdown of this adiabatic prediction. Here $L = 2$, giving the largest mirror ratio among the three cases, $R = 26$, while the launch angle is again $\theta = \pi/4\,\mathrm{rad}$, far above the critical angle $\theta_c = 0.0385\,\mathrm{rad}$. Equation~\eqref{eq:xt} predicts nominal turning points at $x_t = \pm 1.29$. Unlike in Fig.~\ref{fig:collisions}(a), however, these locations do not act as barriers to the longitudinal motion. The trajectory repeatedly crosses both nominal turning points, explores a much larger portion of the channel, and eventually escapes through the left aperture. The large mirror ratio therefore does not guarantee confinement when the adiabatic approximation breaks down.

    The relative variation of the transverse action for the three trajectories is shown in Fig.~\ref{fig:collisions}(d). For the confined trajectory of Fig.~\ref{fig:collisions}(a), $J_{\perp,n}$ remains close to its initial value, with small oscillations throughout the motion. The escaping trajectory of Fig.~\ref{fig:collisions}(b) also exhibits only a small variation of $J_{\perp,n}$ before reaching the aperture, consistent with the adiabatic prediction that a trajectory launched inside the nominal loss cone escapes. By contrast, for the trajectory of Fig.~\ref{fig:collisions}(c), $J_{\perp,n}$ undergoes large variations and eventually differs from its initial value by an amount of order unity. The failure of the nominal turning points in this case is therefore directly associated with the loss of adiabatic invariance. The condition for the validity of the adiabatic approximation is derived below, while the structure of the trapped and escaping regions beyond this regime is examined in Sec.~\ref{sec:phase}.

    Comparing Eq.~\eqref{eq:cone} with Eq.~\eqref{eq:coneB} shows that the billiard and magnetic loss cones have different dependence on their respective mirror ratios. Let $\beta$ denote the compression factor at the nominal turning point, defined as $\beta = h_0/h(x_t)$ for the billiard and $\beta = B_{\mathrm{turn}}/B_0$ for the magnetic mirror. The turning condition is then $\beta\sin\theta = 1$ for the billiard and $\sqrt{\beta}\sin\theta = 1$ for the magnetic mirror. The two loss cones therefore coincide when
    \begin{equation*}
        R_{\rm mag} = R^2.
    \end{equation*}
    Thus, a billiard with geometric mirror ratio $R$ is the analogue of a magnetic mirror with ratio $R^2$ when the systems are compared at matched loss cones.

    Everything from Eq.~\eqref{eq:turning} onwards rests on the adiabatic invariance of $J_\perp$, which holds only if the channel width changes little over one transverse period~\cite{Gardner1959,Wells2007,Neishtadt2019}. We now make this requirement explicit and determine whether the billiard can satisfy it. The transverse period is the time required for the particle to cross the channel and return,
    \begin{equation}
        \label{eq:Tb}
        T_b = \frac{4h}{v_\perp} = \frac{4h}{v\sin\theta}.
    \end{equation}
    while the rate at which the particle sees the half-width change is $\dot h = h'(x)\dot{x} = h'(x)v_\parallel = h'(x)v\cos\theta$. Over one transverse period, the corresponding change in the half-width is approximately $\Delta h \approx \dot h\,T_b$. We therefore define the dimensionless adiabaticity parameter $\varepsilon$ as the fractional change of the half-width during one transverse oscillation,
    \begin{equation}
        \label{eq:epsilon}
        \varepsilon \equiv \frac{\abs*{\Delta h}}{h} \simeq \frac{\abs*{\dot h}}{h}T_b = 4\abs*{h'(x)}\cot\theta.
    \end{equation}
    Adiabaticity requires $\varepsilon \ll 1$. The resulting criterion depends only on the local wall slope and the pitch angle: slowly varying walls and sufficiently large pitch angles favor the conservation of $J_\perp$.


    The dependence on the pitch angle is an important difference from the magnetic mirror. From Eq.~\eqref{eq:Tb}, the transverse bouncing frequency is proportional to $\sin\theta$ and therefore decreases as the pitch angle becomes smaller. By contrast, the cyclotron frequency $\omega_c = qB/m$ is independent of the pitch angle. The adiabatic approximation in the billiard consequently becomes increasingly restrictive as $\theta$ decreases.

    To obtain a sufficient condition for adiabaticity throughout the billiard, we evaluate Eq.~\eqref{eq:epsilon} using the largest wall slope and the smallest pitch angle relevant to the nominal loss-cone boundary. The maximum slope occurs at the inflection points, Eq.~\eqref{eq:maxslope}, while the critical angle satisfies $\sin\theta_c = 1/R$ and hence $\cot\theta_c = \sqrt{R^2 - 1}$. This gives
    \begin{equation}
        \label{eq:validity}
        \frac{L}{h_0} \gg \frac{3\sqrt{3}}{2}\sqrt{R^2 - 1} \approx 2.6R,
    \end{equation}
    where the last approximation applies for $R \gg 1$. Since the local pitch angle increases as the particle approaches a nominal turning point, taking $\theta = \theta_c$ throughout gives the largest value of $\cot\theta$ relevant to the confined trajectories and therefore the most restrictive adiabaticity condition.

    Equation~\eqref{eq:validity} shows that increasing $L$ favors adiabaticity, while Eq.~\eqref{eq:R} shows that it simultaneously decreases the mirror ratio $R$. We can make this competition explicit by eliminating $L$. Isolating $L$ in Eq.~\eqref{eq:R}, we have
    \begin{equation*}
        L = \frac{x_m}{\sqrt{R - 1}}.
    \end{equation*}
    Then, Eq.~\eqref{eq:validity} becomes
    \begin{equation}
        \label{eq:validity-R}
        \frac{x_m}{h_0} \gg \frac{3\sqrt{3}}{2}(R - 1)\sqrt{R + 1}.
    \end{equation}
    Equation~\eqref{eq:validity-R} shows that larger mirror ratios can remain compatible with adiabaticity only if the aperture aspect ratio $x_m/h_0$ is also increased. For $R \gg 1$, the right-hand side scales as $R^{3/2}$, so the largest admissible mirror ratio scales as $(x_m/h_0)^{2/3}$.

    For the values used here, $h_0 = 1$ and $x_m = 10$, the two sides of Eq.~\eqref{eq:validity-R} become comparable at $R \approx 2.9$. From Eq.~\eqref{eq:R}, this corresponds to
    \begin{equation}
        \label{eq:validity-R2}
        L \approx 7.2,
    \end{equation}
    while Eq.~\eqref{eq:cone} gives $\theta_c \approx 0.35\,\mathrm{rad}$. The adiabatic approximation is therefore expected to be reliable only towards the long-profile end of the range studied below. Shortening the profile produces a narrower nominal loss cone, but simultaneously makes the adiabatic approximation less reliable. The strongest confinement is consequently expected near the crossover between these two effects, rather than at the longest profile, where the adiabatic approximation is most accurate but the mirror is weak. Equation~\eqref{eq:cone} should therefore be regarded as a reference prediction rather than an exact escape threshold. The actual boundary between escaping and long-lived initial conditions must be determined from the dynamics and need not coincide with the nominal loss-cone boundary. Because the collision map has a mixed phase space, chaotic trajectories outside the nominal cone can eventually escape, while regular islands and their surrounding structures can support trajectories for much longer times. The resulting boundary is shaped by the invariant structures of the mixed phase space, including invariant curves and cantori~\cite{Mackay1984, Mackay1985, Efthymiopoulos1997}, typical of two--dimensional, area--preserving maps. 

    The analysis above therefore provides an adiabatic reference picture of the billiard, together with a nominal loss cone and a criterion for the validity of the underlying approximation. The actual confinement properties, however, depend on the structure of the full collision map. In the next section, we analyze its phase space and identify the invariant structures associated with regular confinement, chaotic transport, and escape.

    \section{Phase-space structure and confinement}
    \label{sec:phase}

    \begin{figure*}[t!]
        \centering
        \includegraphics[width=\linewidth]{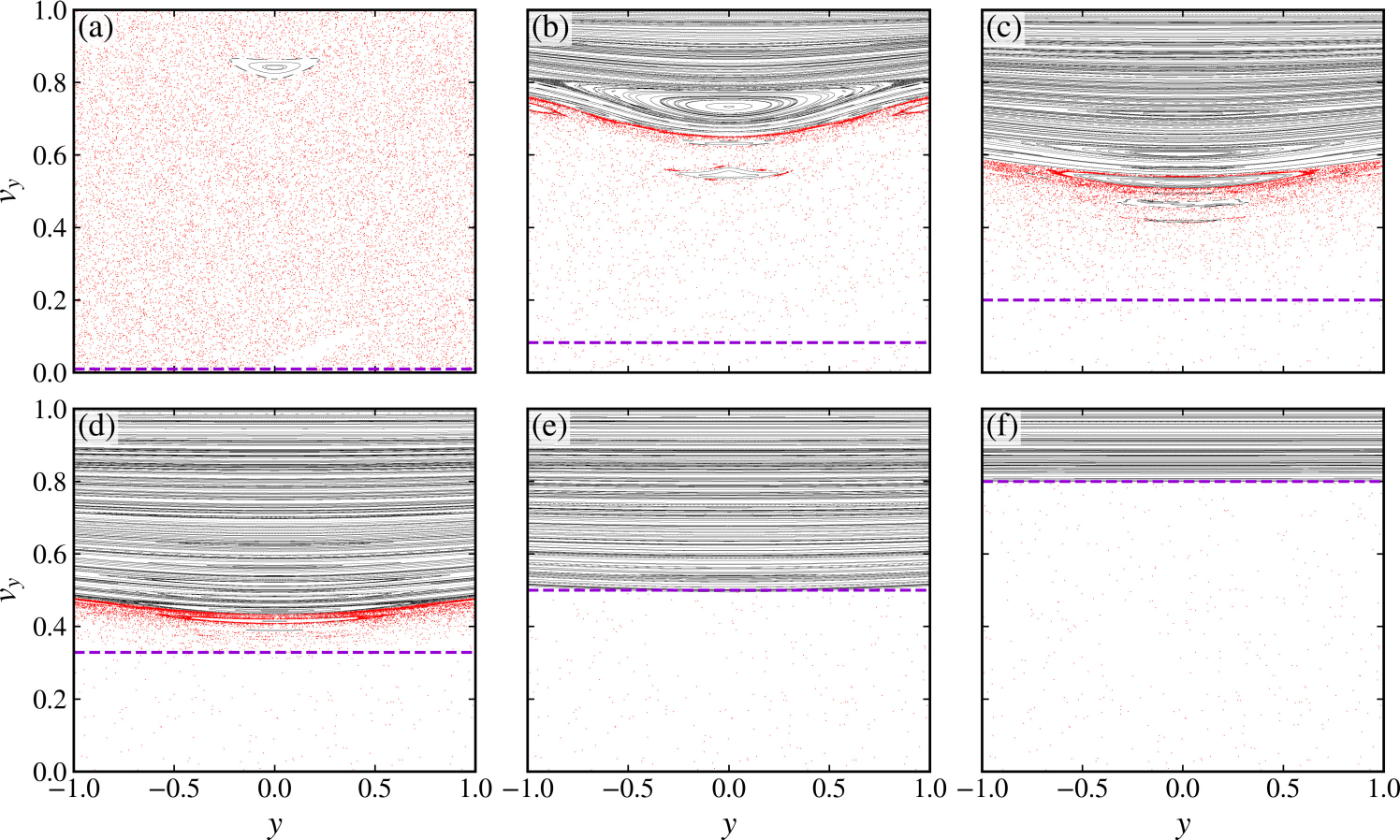}
        \caption{Poincaré section of the billiard, Eq.~\eqref{eq:section}, for $h_0 = 1$, $x_m = 10$ and (a) $L = 1$, (b) $L = 3$, (c) $L = 5$, (d) $L = 7$, (e) $L = 10$ and (f) $L = 20$, for which Eq.~\eqref{eq:R} gives $R = 101$, $12.11$, $5$, $3.04$, $2$ and $1.25$, respectively. In each panel, 500 initial conditions are sampled uniformly from $(y, v_y) \in [-1, 1] \times [0, 1]$ and iterated for $10^4$ crossings. Only $v_y \geq 0$ is shown. Trajectories that remain inside the billiard for $10^4$ crossings are shown in black, while escaping trajectories are depicted in red. The red layer below the black region consists of transient trajectories that spend a finite number of crossings near the trapped region before escaping. The purple dashed line marks the edge of the nominal loss cone, $v_y = 1/R$ from Eq.~\eqref{eq:cone}.}
        \label{fig:phasespaces}
    \end{figure*}

    We now examine how the confinement predicted by the adiabatic approximation is modified by the phase-space structure of the full collision dynamics. The trajectories are generated by the collision map of Sec.~\ref{sec:billiard}, and the phase space is displayed on the Poincaré surface of section (PSS) formed by the midplane crossings~\cite{lichtenberg2013regular},
    \begin{equation}
        \label{eq:section}
        \Sigma = \left\{ x = 0, \quad v_x > 0 \right\},
    \end{equation}
    recording the pair $(y,v_y)$ at each crossing. At fixed speed, the condition $v_x > 0$ gives $v_x = \sqrt{v^2-v_y^2}$, so $\Sigma$ is two dimensional, and $(y,v_y)$ form a pair of canonically conjugate coordinates, as shown in Appendix~\ref{app:section}. The return map on $\Sigma$ preserves $\dd{y}\dd{v_y}$ and acts on the rectangle $\abs*{y} \leq h_0$, $\abs*{v_y} \leq v$. Thus, Lebesgue measure in these variables is invariant, so a uniform sampling of the rectangle corresponds to a uniform sampling of the phase space, and areas measured on the section represent phase-space volumes. On the section $\Sigma$, the midplane loss-cone condition of Eq.~\eqref{eq:cone} therefore corresponds to the horizontal band $\abs*{v_y} < v/R$.

    The half-width is even in $x$, so the transformation $x \to -x$, $v_x \to -v_x$ maps $\mathcal{D}$ onto itself while exchanging the two apertures. Crossings with $v_x < 0$ are therefore related by symmetry to those with $v_x > 0$ recorded on $\Sigma$. Since the walls are static and the reflections are elastic, the dynamics is also invariant under time reversal. Combining time reversal with the reflection $x \to -x$, $v_x \to -v_x$ maps a point $(y,v_y)$ on $\Sigma$ to $(y,-v_y)$. The section is therefore symmetric about $v_y = 0$, and Fig.~\ref{fig:phasespaces} shows only the half with $v_y \geq 0$.

    Figure~\ref{fig:phasespaces} shows the resulting Poincaré sections for six values of the profile length $L$, with black denoting trajectories that remain inside the billiard for $10^4$ crossings and red denoting trajectories that escape. For $L = 1$ and $R = 101$, shown in Fig.~\ref{fig:phasespaces}(a), the billiard has the strongest mirror of the six cases and a nominal loss cone with critical angle $\theta_c = 0.0099\,\mathrm{rad}$. Nevertheless, it confines almost nothing as escaping trajectories fill nearly the entire section, while a single regular island survives near $v_y = 0.85$. The predicted edge of the loss cone, shown as the purple dashed line at $v_y = 0.0099$, lies far below the surviving regular structure. At $L = 3$, shown in Fig.~\ref{fig:phasespaces}(b), a region of invariant curves appears, and this region occupies a larger portion of the section at $L = 5$, shown in Fig.~\ref{fig:phasespaces}(c). These invariant curves reveal the extent to which the transverse action is preserved by the collision map. Since the half-width is $h_0$ at every midplane crossing, conservation of $J_\perp$ implies that $\abs*{v_y}$ remains unchanged from one crossing to the next. The corresponding invariant curves are therefore horizontal. Their bending indicates a systematic variation of $\abs*{v_y}$ between successive crossings and hence a breakdown of the adiabatic invariance of $J_\perp$. In Figs.~\ref{fig:phasespaces}(b) and \ref{fig:phasespaces}(c), only the curves nearest $v_y = v = 1$ are approximately horizontal. The remaining curves bend, with $\abs*{v_y}$ smallest near $y = 0$ and largest near the walls, and the bending becomes stronger as $v_y$ decreases. This trend is consistent with Eq.~\eqref{eq:epsilon}: as $v_y$ decreases, the pitch angle decreases and $\varepsilon$ increases, indicating a progressive loss of adiabaticity. At $v_y = v = 1$, the motion is purely transverse, so $v_\parallel = 0$ and $\varepsilon = 0$. The adiabatic approximation is therefore most accurate near the top of the section and becomes progressively less accurate towards the bottom, precisely where the nominal loss cone is located.

    Additionally, the bending of the invariant curves decreases as $L$ increases. The curves remain clearly bent in Figs.~\ref{fig:phasespaces}(b) and \ref{fig:phasespaces}(c), become less distorted at $L = 7$ in Fig.~\ref{fig:phasespaces}(d), where $R = 3.04$, and are only slightly bent at $L = 10$ in Fig.~\ref{fig:phasespaces}(e), where $R = 2$. In Fig.~\ref{fig:phasespaces}(e), the residual curvature is particularly clear when compared with the horizontal nominal loss-cone boundary. At $L = 20$, shown in Fig.~\ref{fig:phasespaces}(f), where $R = 1.25$, the curves are horizontal on the scale of the figure and the lower boundary of the nonescaping region nearly coincides with the nominal loss-cone boundary. The adiabatic description is therefore progressively recovered as $L$ increases. For the shorter and intermediate profiles, the lower boundary of the nonescaping region depends on both $y$ and $v_y$, whereas the adiabatic loss-cone condition of Eq.~\eqref{eq:cone} depends only on $v_y$. As $L$ increases, this boundary becomes progressively flatter and approaches the nominal loss-cone prediction. Trajectories outside the nominal loss cone can therefore escape when the adiabatic approximation is not sufficiently accurate, while in the long-profile regime the loss-cone criterion provides an increasingly accurate description of the boundary between escaping and nonescaping trajectories.

    Despite this progressive recovery of the adiabatic description, better adiabaticity does not imply better confinement. As $L$ increases, the invariant curves become flatter, but the mirror ratio decreases and the nominal loss cone widens. The nonescaping region is therefore widest at $L = 7$ in Fig.~\ref{fig:phasespaces}(d), rather than at $L = 20$ in Fig.~\ref{fig:phasespaces}(f), where the dynamics is nearly adiabatic but the nominal loss cone occupies most of the section. Its lower boundary descends from Fig.~\ref{fig:phasespaces}(b) through Fig.~\ref{fig:phasespaces}(c) to Fig.~\ref{fig:phasespaces}(d), then rises again in Figs.~\ref{fig:phasespaces}(e) and \ref{fig:phasespaces}(f). The estimate of Eq.~\eqref{eq:validity-R} places the crossover between these competing effects near $L \approx 7.2$ for the geometry used here, close to the value at which the nonescaping region is widest. Immediately below the nonescaping region in Figs.~\ref{fig:phasespaces}(b), \ref{fig:phasespaces}(c), and \ref{fig:phasespaces}(d), a layer of escaping trajectories is dense enough to appear as a continuous band rather than as isolated points. This layer surrounds regular island chains near the boundary of the nonescaping region, where trajectories can remain for many crossings before escaping, a behavior characteristic of stickiness near regular structures in two--dimensional, area--preserving maps~\cite{Efthymiopoulos1997, Contopoulos1971, Meiss1983, Karney1983, Zaslavsky2002,Custodio2011, Carmo2026}.

    \section{Escape statistics}
    \label{sec:escape}

    We now quantify how the competition between adiabaticity and mirror strength affects the fraction of phase space that escapes. The nominal loss cone provides the corresponding adiabatic prediction. On the Poincaré section, the loss cone occupies the band $\abs*{v_y} < v/R$. Since $y \in [-h_0,h_0]$, this band has area
    \begin{equation*}
        A_{\rm lc} = (2h_0)\left(\frac{2v}{R}\right) = \frac{4h_0v}{R}.
    \end{equation*}
    The full section spans $y \in [-h_0,h_0]$ and $v_y \in [-v,v]$, so its area is
    \begin{equation*}
        A_{\Sigma} = (2h_0)(2v) = 4h_0v.
    \end{equation*}
    Because $\dd{y}\dd{v_y}$ is the invariant measure on the section, the fraction of initial conditions predicted to escape is the ratio of these two areas,
    \begin{equation}
        \label{eq:frac}
        f_{\rm ad} = \frac{A_{\rm lc}}{A_{\Sigma}} = \frac{1}{R}.
    \end{equation}
    The prediction is therefore independent of $h_0$ and of the particle speed. We compare it with the escaping fraction $f_{\rm esc}$ obtained directly from the collision dynamics.

    \begin{figure}[t!]
        \centering
        \includegraphics[width=\linewidth]{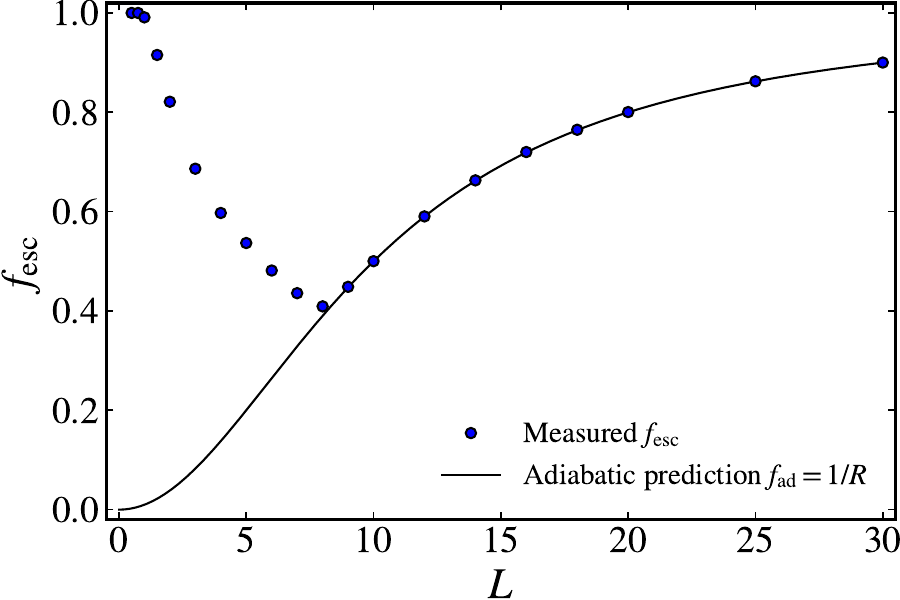}
        \caption{Escaping fraction as a function of the profile length $L$ for $h_0 = 1$ and $x_m = 10$. The blue circles show the measured fraction $f_{\rm esc}$ obtained from $10^6$ initial conditions sampled uniformly on the Poincaré section and evolved for up to $10^6$ collisions. The black solid curve is the adiabatic prediction $f_{\rm ad} = 1/R$ from Eq.~\eqref{eq:frac}.}
        \label{fig:escaping-fraction}
    \end{figure}

    Figure~\ref{fig:escaping-fraction} compares the measured escaping fraction $f_{\rm esc}$, shown by the blue circles, with the adiabatic prediction of Eq.~\eqref{eq:frac}, shown by the black solid curve. For the shortest profiles, the two differ dramatically. As $L \to 0$, the mirror ratio increases and the nominal loss cone narrows, so the adiabatic prediction tends to zero. In contrast, the measured escaping fraction is close to unity for the smallest values of $L$ considered. As $L$ increases, $f_{\rm esc}$ decreases and reaches a minimum near $L \approx 8$, while the adiabatic prediction increases monotonically. Beyond this minimum, the measured fraction rises and rapidly approaches the adiabatic prediction, with the two becoming nearly indistinguishable for the longer profiles. Thus, the strongest disagreement with the adiabatic theory occurs precisely in the short-profile regime, where the nominal mirror is strongest but the adiabatic approximation is least accurate.

    The initial decrease of $f_{\rm esc}$ reflects the progressive recovery of the invariant structures seen in Fig.~\ref{fig:phasespaces}. As $L$ increases, the channel varies more slowly and the transverse action is better conserved, allowing a larger portion of phase space to remain nonescaping despite the simultaneous decrease of the mirror ratio. Beyond the minimum of $f_{\rm esc}$, the dynamics is already sufficiently close to the adiabatic regime that the weakening of the mirror becomes the dominant effect. The nominal loss cone then widens with $L$, and the measured escaping fraction approaches the adiabatic prediction.

    The dependence of the adiabatic prediction on $L$ follows directly from Eq.~\eqref{eq:R}. Substituting the mirror ratio into Eq.~\eqref{eq:frac} gives
    \begin{equation*}
        f_{\rm ad} = \frac{L^2}{L^2+x_m^2},
    \end{equation*}
    which increases monotonically from zero to unity with $L$. The nonmonotonic behavior of $f_{\rm esc}$ therefore results from the competition between the improvement of adiabaticity and the simultaneous weakening of the geometric mirror. Additionally, the minimum of $f_{\rm esc}$ occurs close to the crossover $L \approx 7.2$ estimated from Eqs.~\eqref{eq:validity-R} and \eqref{eq:validity-R2}. Confinement is therefore strongest at an intermediate profile length, where the mirror remains sufficiently strong while the transverse action is sufficiently well preserved. 

    \begin{figure*}[t!]
        \centering
        \includegraphics[width=\linewidth]{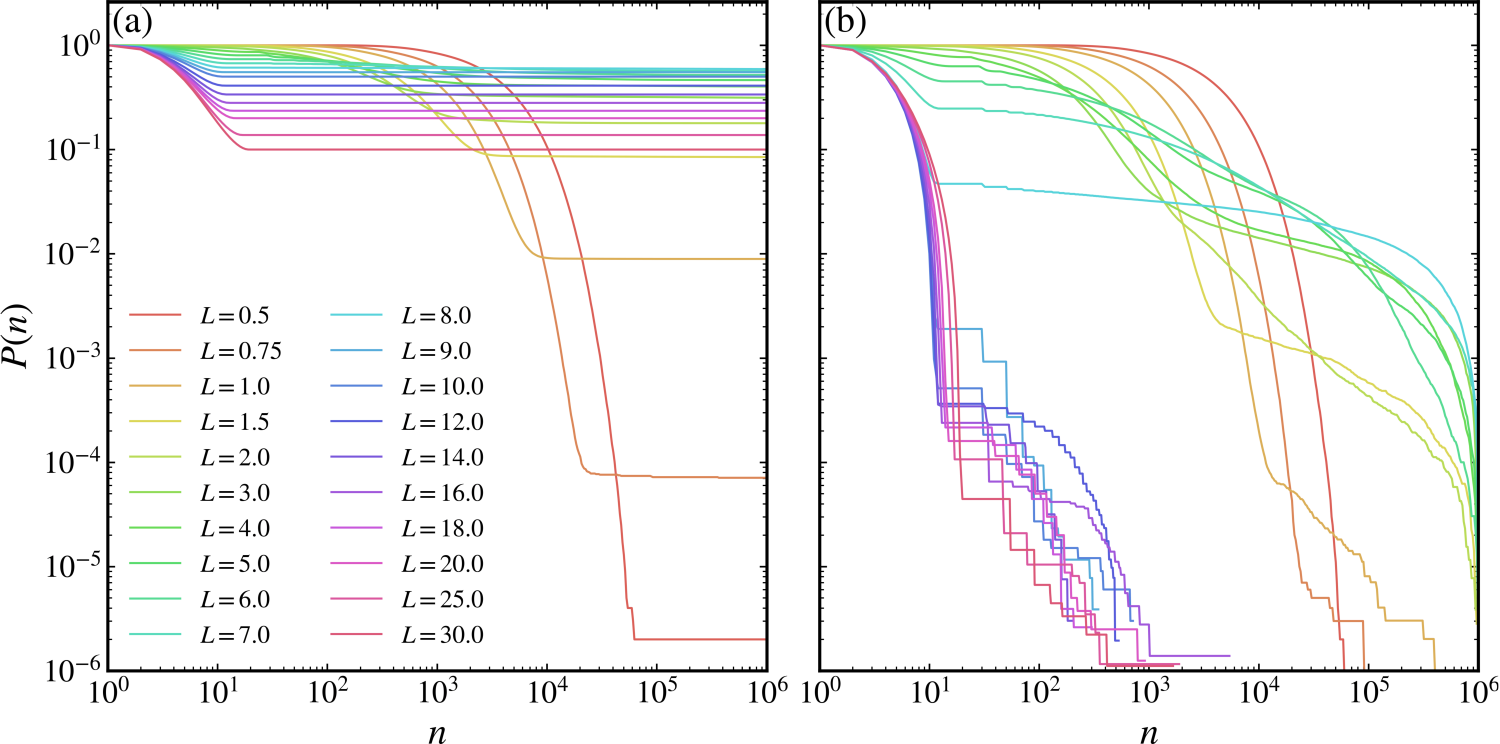}
        \caption{Survival probabilities for $h_0 = 1$ and $x_m = 10$ at different values of the profile length $L$. Each curve is obtained from $10^6$ initial conditions sampled uniformly on the Poincaré section and evolved for up to $n_{\max} = 10^6$ collisions. (a) Survival probability $P(n)$ of the complete ensemble. Trajectories that remain inside the billiard at $n_{\max}$ contribute to the nonzero long-time plateau. (b) Conditional survival probability $P_{\rm esc}(n)$ restricted to trajectories that escape before $n_{\max}$.}
        \label{fig:survival}
    \end{figure*}

    The escaping fraction does not distinguish between trajectories that leave the billiard promptly and those that remain inside for many collisions before escaping. We therefore characterize each escaping trajectory by its escape collision number $n_{\rm esc}$, defined as the number of collisions undergone before reaching either aperture. We first consider the survival probability, commonly used to characterize escape in open dynamical systems~\cite{Dettmann2009,Custodio2011,Altmann2013,Hansen2018,Sales2024,Borin2026,Haerter2026},
    \begin{equation*}
        P(n) = \Pr(n_{\rm esc} \geq n),
    \end{equation*}
    where trajectories that have not escaped by the numerical cutoff $n_{\max} = 10^6$ are retained as survivors. Thus, $P(n)$ gives the fraction of the complete initial ensemble that remains inside the billiard after $n$ collisions.

    Figure~\ref{fig:survival}(a) shows $P(n)$ for the same range of profile lengths considered in Fig.~\ref{fig:escaping-fraction}. The curves generally exhibit an initial decay followed by a plateau whose height depends strongly on $L$. At the numerical cutoff,
    \begin{equation*}
        P(n_{\max}) = 1 - f_{\rm esc},
    \end{equation*}
    so the plateau corresponds to the fraction of trajectories that have not escaped within the observation window. For the shortest profiles, it is nearly zero because almost the entire ensemble escapes. It increases as $L$ enters the intermediate regime, consistently with the growth of the nonescaping region in Fig.~\ref{fig:phasespaces}, and decreases again for long profiles as the nominal loss cone widens.

    The plateau in $P(n)$ masks the statistics of trajectories that actually escape. We therefore also consider the survival probability conditioned on escape within the observation window [Fig.~\ref{fig:survival}(b)],
    \begin{equation*}
        P_{\rm esc}(n) = \Pr\left(n_{\rm esc} \geq n\,\middle|\,n_{\rm esc} < n_{\max}\right).
    \end{equation*}
    By excluding trajectories that remain inside the billiard up to $n_{\max}$, $P_{\rm esc}(n)$ isolates the escape-time statistics of the escaping population. For the shortest profiles, the survival probability remains close to unity for a substantial number of collisions and then decreases comparatively rapidly. As $L$ increases into the intermediate regime, the decay becomes much broader and develops a long-lived tail extending over several decades in $n$. The escaping population is then strongly heterogeneous, with most trajectories escaping on relatively short time scales while a much smaller fraction survives for very long times. For larger $L$, this long-lived tail is strongly suppressed and most escaping trajectories leave within only a few tens of collisions, although a small number survive for longer times. This progression is consistent with the recovery of the adiabatic regime as $L$ increases.

    To quantify the separation between typical and large escape collision numbers, we consider the median and mean of the distribution. The median escape collision number $n_{1/2}$ is defined by
    \begin{equation*}
        P_{\rm esc}(n_{1/2}) = \frac{1}{2},
    \end{equation*}
    while $\langle n_{\rm esc}\rangle$ is the mean escape collision number over trajectories that escape before $n_{\max}$. The median characterizes the time scale of a typical escaping trajectory, whereas the mean is particularly sensitive to the long-lived tail of the distribution. Their ratio therefore measures the extent to which rare long-lived trajectories increase the mean relative to the median.

\begin{figure}[t!]
        \centering
        \includegraphics[width=\linewidth]{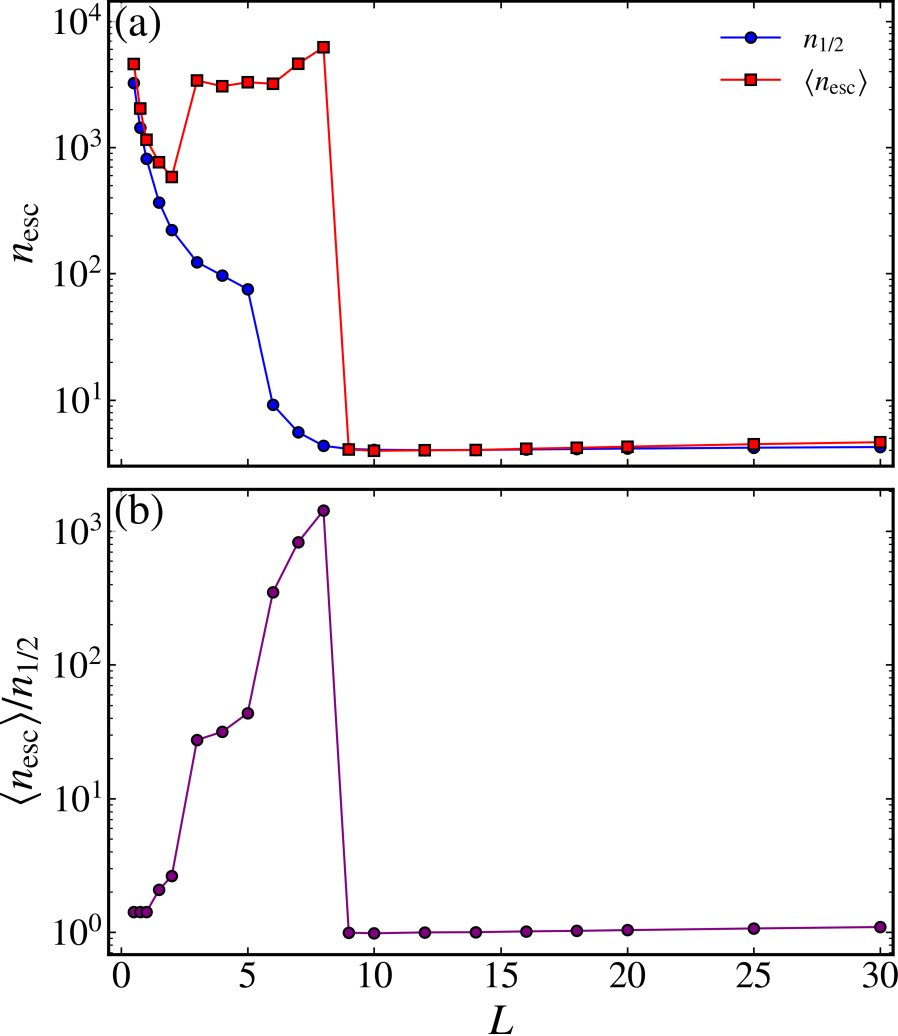}
        \caption{Characteristic escape collision numbers as a function of the profile length $L$ for trajectories that escape before $n_{\max} = 10^6$. (a) Median escape collision number $n_{1/2}$ (in blue/circles) and mean escape collision number $\langle n_{\rm esc}\rangle$ (in red/squares). (b) Ratio $\langle n_{\rm esc}\rangle/n_{1/2}$, which quantifies the separation between the mean and a typical escape collision number. The same ensemble of $10^6$ initial conditions used in Fig.~\ref{fig:survival} is used here.}
        \label{fig:escape-times}
    \end{figure}

    Figure~\ref{fig:escape-times}(a) shows that the median escape collision number decreases continuously with $L$, from several thousand collisions for the shortest profiles to only a few collisions by $L \approx 9$. The mean initially decreases together with the median, but their behavior separates strongly at intermediate $L$. From approximately $L = 3$ to $L = 8$, the median continues to decrease while the mean remains of the order of several thousand collisions and reaches its largest value near $L = 8$. The escaping population in this regime is therefore dominated numerically by trajectories with comparatively small escape collision numbers, while a much smaller population of trajectories with very large escape collision numbers strongly increases the mean. At $L = 9$, the mean drops abruptly to the same scale as the median, and the two remain close for larger $L$.

    This separation is quantified by the ratio $\langle n_{\rm esc}\rangle/n_{1/2}$ shown in Fig.~\ref{fig:escape-times}(b). The ratio is close to unity for the shortest profiles, then increases rapidly with $L$, reaching values of order $10^3$ near $L = 7$ and exceeding $10^3$ at $L = 8$. The large ratio reflects the increasingly strong influence of the long-lived tail on the mean escape collision number. Between $L = 8$ and $L = 9$, the ratio collapses to approximately unity and remains close to one throughout the long-profile regime. The disappearance of this separation coincides with the suppression of the long-lived tail seen in Fig.~\ref{fig:survival}(b).

    The escape dynamics therefore changes sharply between the intermediate and long-profile regimes. At intermediate $L$, the escaping population spans widely separated time scales, whereas for $L \gtrsim 9$ the mean and median become comparable and escape is dominated by short-lived trajectories. This change is consistent with the phase-space evolution of Sec.~\ref{sec:phase}, where increasing $L$ produces a progressively cleaner separation between the nonescaping region and the escaping region associated with the nominal loss cone.

    \section{Escape basins and fractal dimension}
    \label{sec:basins}

\begin{figure*}[t!]
        \centering
        \includegraphics[width=\linewidth]{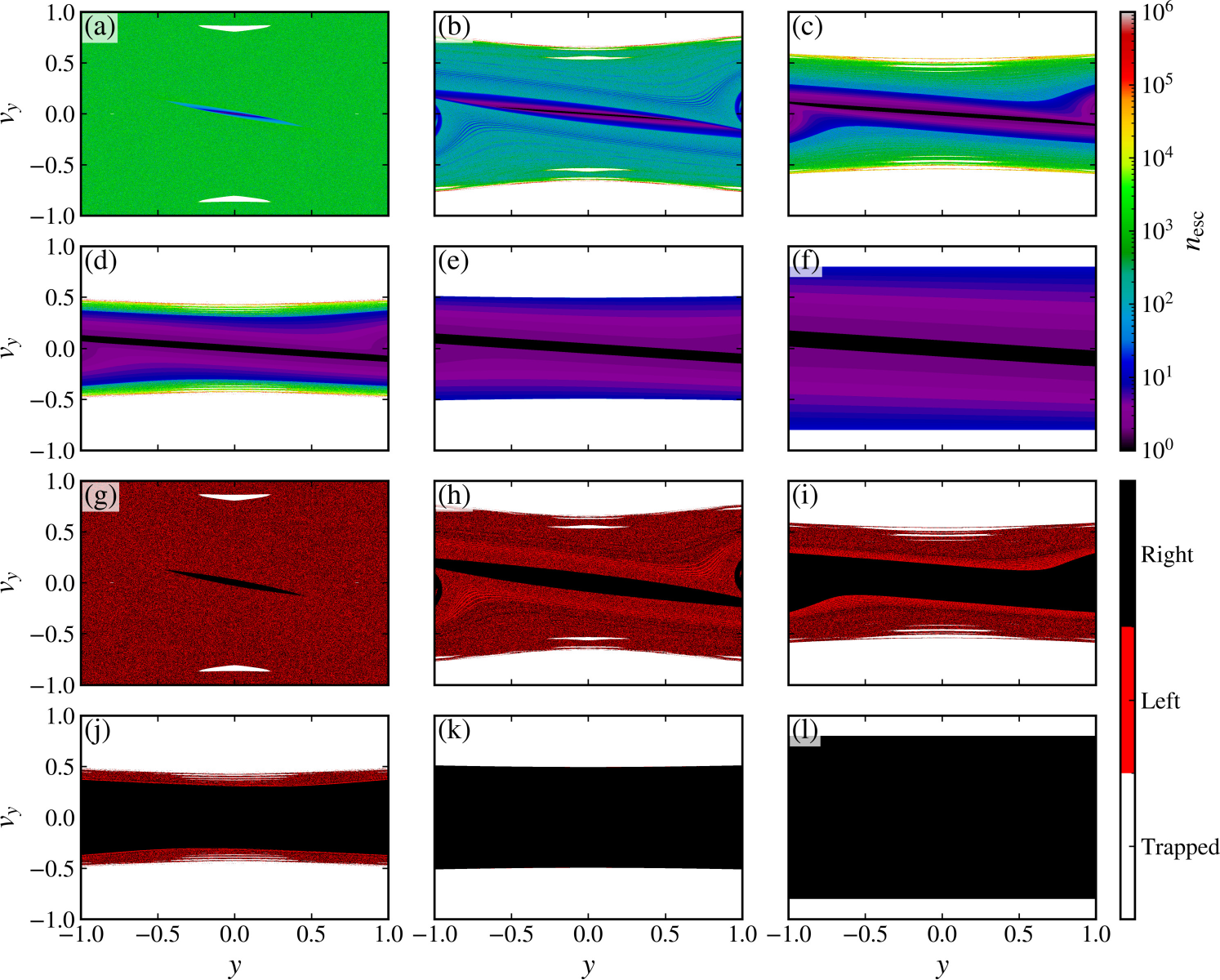}
        \caption{Escape collision number (first two rows) and escape basins (last two rows) on the Poincaré section $\Sigma$ for $h_0 = 1$, $x_m = 10$ and (a,g) $L = 1$, (b,h) $L = 3$, (c,i) $L = 5$, (d,j) $L = 7$, (e,k) $L = 10$ and (f,l) $L = 20$. The upper two rows show the escape collision number $n_{\rm esc}$ on a logarithmic color scale. Initial conditions that remain inside the billiard up to the cutoff $n_{\max}$ are shown in white. The lower two rows show the corresponding escape basins, with white denoting trajectories that remain inside up to $n_{\max}$, red denoting escape through the left aperture and black denoting escape through the right aperture.}
        \label{fig:basins}
    \end{figure*}

    The survival probabilities of Sec.~\ref{sec:escape} characterize the escape collision numbers statistically, but they do not show where trajectories with different escape collision numbers originate in phase space or which aperture they ultimately reach. We therefore return to the Poincaré section $\Sigma$ and classify each initial condition according to its escape collision number and escape direction. Figure~\ref{fig:basins} shows both quantities for the same six values of $L$ considered in Fig.~\ref{fig:phasespaces}. For $L = 1$, shown in Figs.~\ref{fig:basins}(a) and \ref{fig:basins}(g), almost the entire section consists of escaping trajectories, apart from the small regular islands near $\abs*{v_y} \approx 0.85$. The escape collision numbers in Fig.~\ref{fig:basins}(a) are comparatively homogeneous over most of the section, while a narrow diagonal region of smaller escape collision numbers crosses its central part. The corresponding escape basins in Fig.~\ref{fig:basins}(g) have a much finer structure, with trajectories escaping through the two apertures strongly intermingled over a large portion of the section and a pronounced diagonal region associated predominantly with right escape. Thus, even when the spatial distribution of escape collision numbers is relatively simple on large scales, the final escape direction can depend sensitively on the initial condition.

    For $L = 3$ and $L = 5$, shown in Figs.~\ref{fig:basins}(b), \ref{fig:basins}(h), \ref{fig:basins}(c), and \ref{fig:basins}(i), the nonescaping regions occupy a substantial portion of the section, while the escaping regions develop a pronounced dependence on both $y$ and $v_y$. The left- and right-escape basins remain finely intermingled and organize along the curved phase-space structures. The escape collision numbers also vary strongly with the initial condition, with trajectories having the largest escape collision numbers concentrated near the boundaries of the nonescaping regions and smaller escape collision numbers occurring farther inside the escaping region. Thus, trajectories approaching the nonescaping regions can undergo increasingly long transient motion before escape.

    For $L = 7$, shown in Figs.~\ref{fig:basins}(d) and \ref{fig:basins}(j), the nonescaping region occupies its largest portion among the cases shown, consistently with the minimum of the escaping fraction in Fig.~\ref{fig:escaping-fraction}. The escaping trajectories occupy a central region around $v_y = 0$ whose boundaries remain visibly curved. Escape is fastest near the center of this region and becomes progressively slower towards the nonescaping region. The right-escape basin dominates most of the central region, while the left-escape basin is concentrated near its upper and lower boundaries, where the two escape outcomes remain finely intermingled.

    For $L = 10$ and $L = 20$, shown in Figs.~\ref{fig:basins}(e), \ref{fig:basins}(k), \ref{fig:basins}(f), and \ref{fig:basins}(l), the structure becomes considerably simpler. The central escaping region widens as the nonescaping region recedes towards larger $\abs*{v_y}$, while its boundaries become progressively flatter and depend less strongly on the initial position $y$. At $L = 10$, escape through the right aperture occupies almost the entire central region, and the fine intermingling between the two escape basins observed at smaller $L$ has nearly disappeared. The escape collision number is also organized predominantly by $v_y$, increasing from the center towards the boundary of the nonescaping region. At $L = 20$, this organization is even more pronounced. The boundaries are nearly horizontal and the central region is entirely dominated by escape through the right aperture. These features are consistent with the recovery of the adiabatic description, for which the midplane loss-cone condition of Eq.~\eqref{eq:cone} depends only on $v_y$.

    The predominance of the right-escape basin follows directly from the definition of the Poincaré section. Since $\Sigma$ contains only crossings with $v_x > 0$, every initial condition represented in Fig.~\ref{fig:basins} initially moves towards the right aperture. In the adiabatic regime, a trajectory inside the loss cone has no turning point before reaching that aperture and therefore escapes to the right. Escape through the left aperture requires the longitudinal motion to reverse before the trajectory eventually escapes. The imbalance between the two escape basins, particularly evident at large $L$, therefore results from the choice of $v_x > 0$ in the definition of $\Sigma$, rather than from any asymmetry of the billiard.

    At $L = 1$, the left- and right-escape basins are finely intermingled over much of the section even though the escape collision numbers are comparatively homogeneous over large regions. At intermediate $L$, large escape collision numbers become concentrated near the boundaries of the nonescaping regions, while the two escape outcomes remain strongly intermingled in parts of the escaping region. Sensitivity of the escape destination to the initial condition therefore does not by itself imply a large escape collision number. Additionally, the fine intermingling of the left- and right-escape basins in Fig.~\ref{fig:basins} indicates a strong sensitivity of the escape direction to the initial condition, a characteristic feature of fractal basin boundaries in open dynamical systems~\cite{Aguirre2009, Portela2007, Kroetz2008}. We quantify this sensitivity using the uncertainty fraction method~\cite{Grebogi1983,McDonald1985,Grebogi1987}, which measures how the probability of changing the final escape direction depends on the magnitude of a perturbation to the initial condition.

    Consider an initial condition $\vb{z}_0 = (y_0,v_{y,0})$ on the Poincaré section $\Sigma$. We evolve the corresponding trajectory until it escapes and classify it according to the aperture through which escape occurs. We then choose a second initial condition $\vb{z}_\varepsilon$ at Euclidean distance $\varepsilon$ from $\vb{z}_0$ in a uniformly random direction and determine its escape direction. Reference points within a distance $\varepsilon$ of the boundary of the sampled domain are rejected so that the perturbed initial condition remains inside the domain. If $\vb{z}_0$ and $\vb{z}_\varepsilon$ escape through different apertures, $\vb{z}_0$ is classified as $\varepsilon$-uncertain. Pairs for which either trajectory does not escape within the maximum number of collisions are excluded, since we are interested specifically in the boundary separating the left- and right-escape basins. Repeating this procedure for a large number of randomly sampled initial conditions gives the uncertainty fraction
    \begin{equation*}
        f(\varepsilon) = \frac{N_{\rm unc}(\varepsilon)}{N_{\rm esc}(\varepsilon)},
    \end{equation*}
    where $N_{\rm unc}(\varepsilon)$ is the number of $\varepsilon$-uncertain pairs and $N_{\rm esc}(\varepsilon)$ is the number of usable pairs for which both trajectories can be assigned to one of the two escape basins. Thus, $f(\varepsilon)$ measures the sensitivity of the final escape direction without including the boundary between escaping and nonescaping motion.

    For a fractal basin boundary, the uncertainty fraction is expected to scale with the perturbation size as~\cite{Grebogi1983,McDonald1985,Grebogi1987}
    \begin{equation}
        \label{eq:uncertainty-scaling}
        f(\varepsilon) \sim \varepsilon^\alpha,
    \end{equation}
    where $\alpha$ is the uncertainty exponent. If the initial-condition space has dimension $D$ and the basin boundary has capacity dimension $d$, the volume of points lying within a distance $\varepsilon$ of the boundary scales as $\varepsilon^{D-d}$. Comparing this scaling with Eq.~\eqref{eq:uncertainty-scaling} gives~\cite{McDonald1985}
    \begin{equation*}
        \alpha = D - d.
    \end{equation*}
    Since the Poincaré section is two dimensional, $D = 2$, and therefore
    \begin{equation*}
        d = 2 - \alpha.
    \end{equation*}
    A smooth one-dimensional boundary has $\alpha = 1$ and $d = 1$, whereas $0 < \alpha < 1$ corresponds to a fractal boundary. As $\alpha$ decreases, the boundary dimension approaches that of the full section, and in the limit $\alpha \to 0$ one has $d \to 2$.

    For each value of $L$, we evaluate $f(\varepsilon)$ over logarithmically spaced values of $\varepsilon$ using $10^4$ usable pairs of initial conditions for each $\varepsilon$. We determine $\alpha$ from a linear fit of $\log f(\varepsilon)$ against $\log\varepsilon$. Trajectories are evolved for up to $n_{\max} = 10^5$ collisions, and the calculation is repeated for ten independent realizations of the initial-condition ensemble. The reported exponent is the mean of the ten fitted values, while its uncertainty is given by their standard deviation.

    \begin{figure}[t!]
        \centering
        \includegraphics[width=\linewidth]{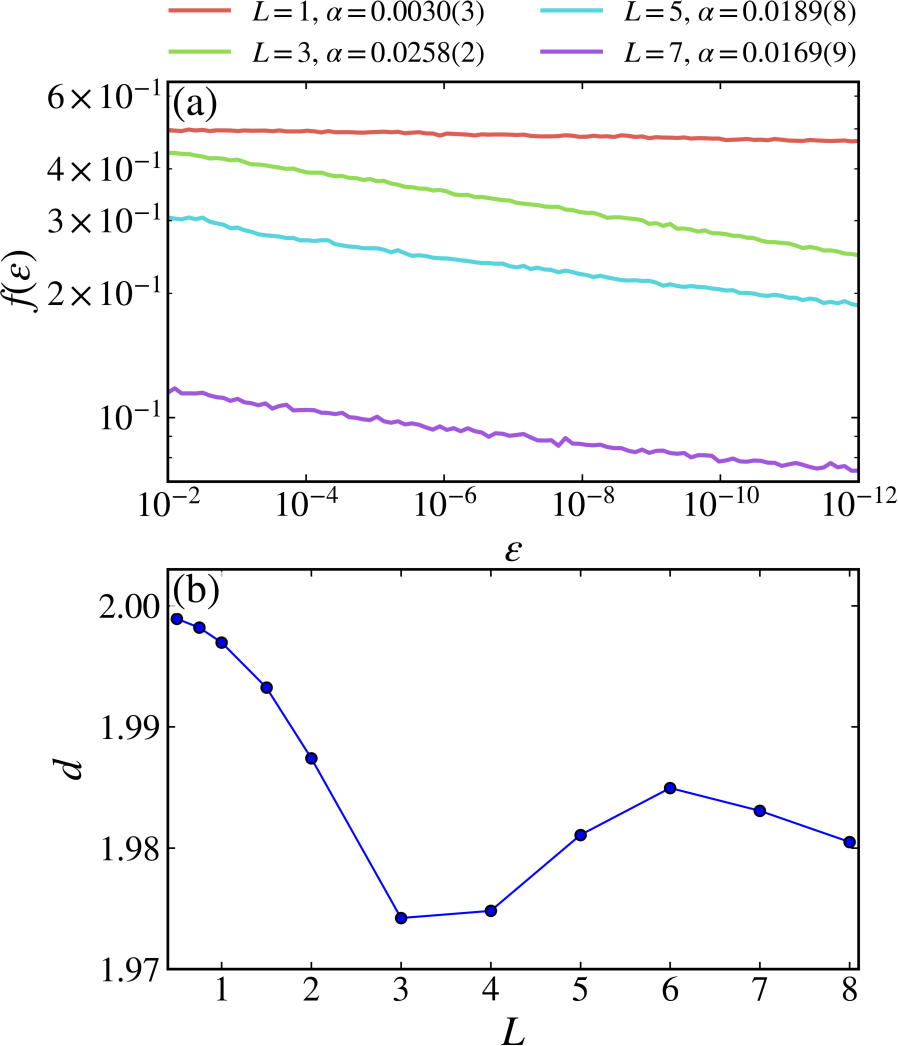}
        \caption{Uncertainty analysis of the left- and right-escape basins for $h_0 = 1$ and $x_m = 10$. (a) Uncertainty fraction $f(\varepsilon)$ as a function of the perturbation size $\varepsilon$ for representative values of the profile length $L$. The curves are obtained using $10^4$ usable pairs of initial conditions for each $\varepsilon$ and a maximum of $n_{\max} = 10^5$ collisions. The values of $\alpha$ shown in the legend are the means over ten independent realizations, with the numbers in parentheses indicating the standard deviation in the last digit. (b) Capacity dimension $d = 2 - \alpha$ of the boundary separating the left- and right-escape basins as a function of $L$.}
        \label{fig:uncertainty}
    \end{figure}

    Figure~\ref{fig:uncertainty}(a) shows that the uncertainty fraction decreases only very slowly as the perturbation size is reduced and the fitted exponents are correspondingly small. For example, we obtain $\alpha = 0.0030(3)$ at $L = 1$, $\alpha = 0.0258(2)$ at $L = 3$, $\alpha = 0.0189(8)$ at $L = 5$, and $\alpha = 0.0169(9)$ at $L = 7$. These values correspond to boundary dimensions $d = 1.9970$, $1.9742$, $1.9811$, and $1.9831$, respectively. Thus, throughout the range considered, the boundary separating the two escape basins has a dimension very close to that of the two-dimensional Poincaré section. The escape direction therefore remains highly sensitive to small perturbations of the initial condition.

    The boundary dimension $d$ depends nonmonotonically on $L$, as shown in Fig.~\ref{fig:uncertainty}(b). Starting from values very close to $d = 2$ for the shortest profiles, it decreases as $L$ increases and reaches its minimum near $L = 3$. It then increases again, reaching a local maximum near $L = 6$, before decreasing towards $L = 8$. Despite these variations, $d$ remains above approximately $1.97$ throughout the interval. The changes with $L$ therefore modify the degree of fractality quantitatively without removing the nearly space-filling character of the basin boundary. Because the classification of escaping and nonescaping trajectories depends on the finite collision cutoff, we also repeated the calculation using a larger value of $n_{\max}$. The estimated dimensions change slightly, as expected when very long-lived trajectories are reclassified, but the dependence on $L$ and the conclusion that the basin boundary remains close to two dimensional are unchanged.

    \section{Conclusions}
    \label{sec:conclusions}

    In this work, we have introduced a stationary, field-free billiard that reproduces the adiabatic confinement mechanism of a magnetic mirror through the geometry of its boundary. The fast gyromotion of a charged particle in the magnetic mirror is replaced by rapid transverse bouncing between the walls, while the slowly varying magnetic field strength is replaced by the longitudinal variation of the channel width. We have shown that the transverse action $J_\perp$ plays the role of the magnetic moment and generates an effective barrier for the longitudinal motion. Approximate conservation of $J_\perp$ leads to the nominal turning point and loss cone of the billiard. The quantitative correspondence between the effective potentials is given by $B(x)/B_0 \longleftrightarrow [h_0/h(x)]^2$, and the loss cones of the two systems coincide when their mirror ratios satisfy $R_{\rm mag} = R^2$.

    We have found that the efficiency of the geometric mirror is controlled by a competition between mirror strength and adiabaticity. Decreasing the profile length increases the geometric mirror ratio and narrows the nominal loss cone, but at the same time makes the channel vary more rapidly and reduces the accuracy of the adiabatic approximation. Increasing the profile length has the opposite effect, improving adiabaticity while weakening the mirror. For the geometry considered here, the adiabaticity estimate gives a crossover near $L \approx 7.2$. The escaping fraction is strongly nonmonotonic and reaches its minimum near $L \approx 7$, showing that the strongest confinement occurs at an intermediate profile length rather than in either the strongest-mirror or most-adiabatic limits.

    We have shown using Poincaré sections how this competition is organized in phase space. For short profiles, the transverse action is poorly conserved and the nominal loss-cone criterion fails to describe the dynamics, with almost the entire section open to escape despite the very narrow nominal loss cone. As $L$ increases, regular invariant structures develop and the nonescaping region grows, reaching its largest extent in the intermediate regime. For still larger $L$, the invariant curves become increasingly horizontal and the lower boundary of the nonescaping region approaches the nominal loss-cone prediction, but the simultaneous weakening of the mirror widens the loss cone and increases the escaping fraction. The loss-cone criterion therefore becomes accurate only when the transverse action is sufficiently well conserved, while outside this regime the full mixed phase-space structure determines confinement and escape.

    We have also shown that the intermediate regime supporting the strongest confinement exhibits particularly heterogeneous transient dynamics. The conditional survival probabilities develop broad long-time tails, and the mean escape collision number becomes orders of magnitude larger than the median. Near $L \approx 8$, their ratio exceeds $10^3$, showing that a relatively small population of trajectories with very large escape collision numbers strongly increases the mean. The largest escape collision numbers are concentrated near the boundaries of the nonescaping regions and the regular structures of the mixed phase space. For larger $L$, the long-lived tail is strongly suppressed, the mean and median become comparable, and escape is dominated by trajectories with small escape collision numbers.

    Finally, we have found that the escape destination can remain highly sensitive to the initial condition even when the escape collision number is comparatively small. The left- and right-escape basins are strongly intermingled over a substantial range of profile lengths. The uncertainty analysis gives very small uncertainty exponents, showing that reducing the uncertainty in the initial condition produces only a weak reduction in the uncertainty of the final escape direction. Thus, even with increasingly precise knowledge of the initial condition, it remains difficult to predict through which aperture a trajectory will escape. The corresponding basin-boundary dimensions remain above approximately $1.97$ throughout the range investigated, consistent with the nearly space-filling structure of the boundary. We have also verified that increasing the maximum collision number changes the estimated dimensions only slightly and does not alter their dependence on the profile length.

    Taken together, these results show that adiabatic confinement can be generated by boundary geometry alone, without a magnetic field or a time-dependent boundary. The billiard considered here provides a simple setting in which an adiabatic invariant, mixed Hamiltonian dynamics, confinement, and escape can be studied within the same system. For the half-width profile considered here, increasing the geometric mirror ratio does not necessarily improve confinement, since the associated change in the profile can also reduce the accuracy of the adiabatic approximation. Whether the same competition persists for other channel geometries remains an open question. A natural extension is to consider time-dependent walls, where the geometric confinement mechanism studied here would coexist with the energy exchange and additional dynamical effects introduced by a moving boundary.

    \section*{Code availability}

    The source code to reproduce the results reported in this paper is freely available in the GitHub repository in Ref.~\footnote{\href{https://github.com/mrolims-publications/adiabatic-confinement-billiard}{https://github.com/mrolims-publications/adiabatic-confinement-billiard}}.

    \section*{Acknowledgments}

    M.R.S.~acknowledges financial support from the São Paulo Research Foundation (FAPESP, Brazil) under Grant No.~2023/08698-9. L.C.S.~acknowledges FAPESP under Grant No.~2023/16146-6. E.D.L.~acknowledges FAPESP under Grant No.~2025/14544-0 and the National Council for Scientific and Technological Development (CNPq, Brazil) under Grant No.~304398/2023-3. I.L.C.~acknowledges FAPESP under Grant No.~2024/05700-5 and CNPq under Grant No.~304616/2021-4. R.L.V.~acknowledges partial financial support from CNPq under Grant Nos.~403120/2021-7, 443575/2024-0, and 301019/2019-3, and from the Coordination of Superior Level Staff Improvement (CAPES, Brazil) under Grant No.~88881.895032/2023-01. This research was also supported by INCT-NeuroComp (CNPq Grant No.~408389/2024-9). The computational simulations in this research were supported by resources supplied by the Center for Scientific Computing (NCC/GridUNESP) of the São Paulo State University (UNESP) and by the \href{http://portal.if.usp.br/controle/}{Oscillations Control Group} of the University of São Paulo.

    \appendix

    \section{Canonical structure of the midplane section}
    \label{app:section}

    The free motion inside $\mathcal{D}$ is generated by the Hamiltonian
    \begin{equation*}
        H = \frac{p_x^2 + p_y^2}{2m},
    \end{equation*}
    and preserves the symplectic form $\omega = \dd{x}\wedge\dd{p_x} + \dd{y}\wedge\dd{p_y}$. Specular reflection at the boundary preserves the Hamiltonian structure of the billiard dynamics, so the resulting Poincaré return map is symplectic.

    On a fixed energy shell, $p_x$ is determined by $p_y$ through
    \begin{equation*}
        p_x = \pm\sqrt{2mE - p_y^2},
    \end{equation*}
    and the condition $v_x > 0$ in Eq.~\eqref{eq:section} selects the positive branch. The section $\Sigma$ is defined by $x = 0$, so $\dd{x} = 0$ on its tangent space and the restriction of the symplectic form becomes
    \begin{equation*}
        \omega\big|_\Sigma = \dd{y}\wedge\dd{p_y}.
    \end{equation*}
    The return map on $\Sigma$ therefore preserves the area element $\dd{y}\dd{p_y}$. With $m = 1$, one has $p_y = v_y$, so the invariant area element is $\dd{y}\dd{v_y}$. The variables $(y,v_y)$ can therefore be used as canonical coordinates on the section, which occupies the rectangle $\abs*{y} \leq h_0$, $\abs*{v_y} \leq v$. Uniform sampling in these variables consequently corresponds to uniform sampling with respect to the invariant measure on the Poincaré section.

    A more usual choice for billiards is the collision map in Birkhoff coordinates $(s,\cos\alpha)$, where $s$ is the arclength along the boundary and $\alpha$ is the angle between the outgoing velocity and the tangent. In these coordinates, the map preserves $\dd{s}\wedge\dd{\cos\alpha}$. For the present geometry, however, the midplane section is more convenient. The boundary consists of two disconnected arcs, so the arclength coordinate requires an additional branch index, and the relation
    \begin{equation*}
        s(x) = \int^x \sqrt{1 + h'(x')^2}\,\dd{x'}
    \end{equation*}
    must in general be evaluated numerically. The midplane section avoids both complications and has the additional advantage that the nominal loss cone appears as a horizontal band in $(y,v_y)$.


%

\end{document}